\documentclass[a4paper,twocolumn,british,authoryear,superscriptaddress]{revtex4}
\usepackage[T1]{fontenc}
\usepackage[latin9]{inputenc}
\usepackage{color}
\usepackage{babel}
\usepackage{array}
\usepackage{longtable}
\usepackage{amsmath}
\usepackage{graphicx}
\usepackage[unicode=true,
 bookmarks=false,
 breaklinks=false,pdfborder={0 0 1},backref=false,colorlinks=true]
 {hyperref}
\hypersetup{
 hidelinks,hidelinks,citecolor=blue,hidelinks,citecolor=blue}

\makeatletter

\pdfpageheight\paperheight
\pdfpagewidth\paperwidth

\providecommand{\tabularnewline}{\\}

\@ifundefined{textcolor}{}
{%
 \definecolor{BLACK}{gray}{0}
 \definecolor{WHITE}{gray}{1}
 \definecolor{RED}{rgb}{1,0,0}
 \definecolor{GREEN}{rgb}{0,1,0}
 \definecolor{BLUE}{rgb}{0,0,1}
 \definecolor{CYAN}{cmyk}{1,0,0,0}
 \definecolor{MAGENTA}{cmyk}{0,1,0,0}
 \definecolor{YELLOW}{cmyk}{0,0,1,0}
}

\usepackage{babel}
\usepackage{array}
\usepackage{textcomp}

\makeatother

\begin{document}

\title{Computational prediction of replication sites in DNA sequences using
complex number representation}

\author{Shubham Kundal}

\affiliation{Department of Electrical Engineering, Indian Institute of Technology
(IIT) Delhi, Hauz Khas, New Delhi - 110016, India.}

\author{Raunak Lohiya}

\affiliation{Department of Mathematics, Indian Institute of Technology (IIT) Delhi,
Hauz Khas, New Delhi - 110016, India.}

\author{Hritik Bansal\footnote[2]{Equal Contribution}}

\affiliation{Department of Electrical Engineering, Indian Institute of Technology
(IIT) Delhi, Hauz Khas, New Delhi - 110016, India.}

\author{Shreya Johri\footnotemark[2]}

\affiliation{Department of Biochemical Engineering and Biotechnology, Indian Institute
of Technology (IIT) Delhi, Hauz Khas, New Delhi - 110016, India.}

\author{Varuni Sarwal\footnotemark[2]}

\affiliation{Department of Biochemical Engineering and Biotechnology, Indian Institute
of Technology (IIT) Delhi, Hauz Khas, New Delhi - 110016, India.}

\author{Kushal Shah}
\email{kushals@iiserb.ac.in}

\selectlanguage{british}%

\affiliation{Department of Electrical Engineering and Computer Science, Indian
Institute of Science Education and Research (IISER), Bhopal - 462066,
Madhya Pradesh, India.}
\begin{abstract}
Computational prediction of origin of replication (ORI) has been of
great interest in bioinformatics and several methods including GC-skew, 
auto-correlation etc. have been explored in the past.
In this paper, we have extended the auto-correlation method to predict
ORI location with much higher resolution for prokaryotes and eukaryotes,
which can be very helpful in experimental validation of the computational
predictions. The proposed complex correlation method (iCorr) converts
the genome sequence into a sequence of complex numbers by mapping
the nucleotides to $\left\{ +1,-1,+i,-i\right\} $ instead of $\left\{ +1,-1\right\} $
used in the auto-correlation method (here, $i$ is square root of
$-1$). Thus, the iCorr method exploits the complete spatial information
about the positions of all the four nucleotides unlike the earlier
auto-correlation method which uses the positional information of only
one nucleotide. Also, the earlier auto-correlation method required
visual inspection of the obtained graphs to identify the location
of origin of replication. The proposed iCorr method does away with
this need and is able to identify the origin location simply by picking
the peak in the iCorr graph. 
\end{abstract}

\keywords{Signal processing, Complexity measures, Correlation and regression
analysis.}
\maketitle

\section{Introduction}

\begin{figure*}
\centering{}\includegraphics[width=1\textwidth]{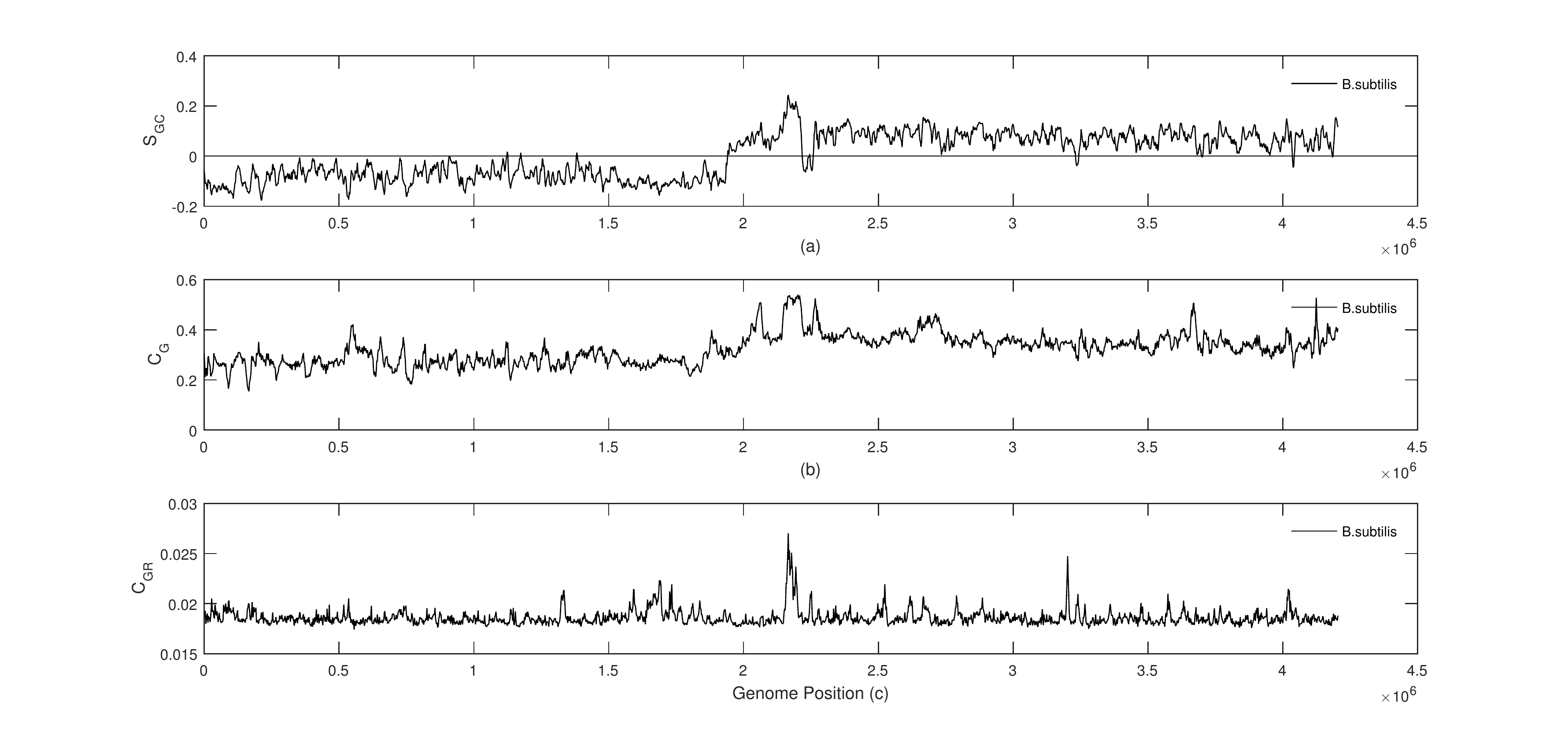} \caption{Plot of the (a) GC skew $\left(S_{GC}\right)$, (b) gCorr $\left(C_{G}\right)$
and (c) iCorr $\left(C_{GR}\right)$ values for \emph{B. subtilis
}subsp. subtilis str. 168 (NC\_000964). As can be seen, for \emph{B.
subtilis}, all three methods predict the TER location at nearby genome
positions. The window size is 10000 and shift size is 2000. \label{fig:01}}
\end{figure*}

DNA replication is a complex biological process by which the genome/chromosome
of an organism creates a copy of itself during cell division. The
segment of DNA sequence where the process of replication initiates is called origin of replication (ORI). 
The ability to computationally predict ORI location is important
to understand the statistical features in a DNA sequence and in the
future, could also provide information for the development of new
drugs for treatment of diseases.

Prokaryotic organisms are usually found to have a single origin of replication
from where two replication forks transmit in contrary directions \cite{Marians1992,Mott2007,Rocha1999}.
More evolved organisms are found to contain multiple sites from which
replication initiates and this helps to speed up the process \cite{Kelman2004,Nasheuer2002}.
Experimental detection of ORI locations is very challenging and so
far has been completed only for a very few archaea, eubacteria and
eukaryotic genomes \cite{Sernova2008}. Here, computational prediction
can play a significant role by considerably reducing the search space
which can save a large amount of experimental time, effort and resources. Computational
prediction of ORI rests on the general hypothesis that the origin
location and its flanking regions have different statistical properties
as compared to rest of the genome. Motivation for this hypothesis
comes from the fact that the replication process of the leading and lagging
strands takes place through a slightly different set of proteins which
can leave certain statistical signatures at the origin location \cite{Lobry1996a,Lobry1996b}.

Different computational methods have been earlier developed to predict origin
of replication in DNA sequences including GC-skew \cite{Lobry1996a,Lobry1996b,Mrazek1998,Touchon2005},
Z-curve \cite{Zhang2005}, CGC Skew \cite{Grigoriev1998}, AT excursion
\cite{Chew2007}, Shannon entropy \cite{Schneider1997,Schneider2010,Shannon1948},
wavelet approach \cite{Song2003}, auto-correlation based measure
\cite{Shah2012}, correlated entropy measure (CEM) \cite{Parikh2015}, GC
profile \cite{Li2014} and few others. All methods use the fundamental
property of identifying differences in statistical properties in the
upstream and downstream side of replication origin to account for mutational
pressures developed in the opening and ending strands of ORI \cite{Lobry2002,Mackiewicz2004}.

In the GC-skew \cite{Mrazek1998} and auto-correlation method \cite{Shah2012}, the entire
genome is divided into overlapping segments/windows and the value
of a certain statistical measure is calculated for each window. For bacterial
genomes, usually the window size is chosen to be around one-hundredth
of the genome size and two consecutive windows have an overlap of
four-fifths of the window size. So, only one-fifth of the genome sequence
is changed per window which helps to reduce the noise produced by sharp
variations of correlation measure in adjacent windows. 

In the GC-skew
method, the number of G and C nucleotides is counted for each segment/window
and the GC-skew value, 
\begin{equation}
S_{GC}=\frac{C-G}{C+G}\label{eq:GC-Skew}
\end{equation}
is plotted against the window number. An ORI (or TER) is then predicted
to be present at the location where the GC-skew value crosses the
zero line from above (or below). The auto-correlation method  (henceforth, called gCorr) goes
a step further and uses the positional information of the G nucleotides
in each window and hence, is informationally richer than the GC-skew
method. Predictions of the gCorr method for ORI location of chromosome 1 and 10 of \emph{P. falciparum} have been recently experimentally verified \cite{Agarwal2017}. It has also been shown earlier that variations of the auto-correlation method
are able to predict the origin location of several
more genomes as compared to the GC-skew method \cite{Parikh2015}.
However, in the auto-correlation method, currently there is no clear
way to differentiate between ORI/TER and the predicted location could
be either of the two for prokaryotic genomes. However, eukaryotic
genomes are linear and do not have a separate TER location. Hence,
the prediction of auto-correlation method invariably corresponds to
the ORI location in eukaryotes.

\begin{figure*}[!tpb]
\centering{}\includegraphics[width=1\textwidth]{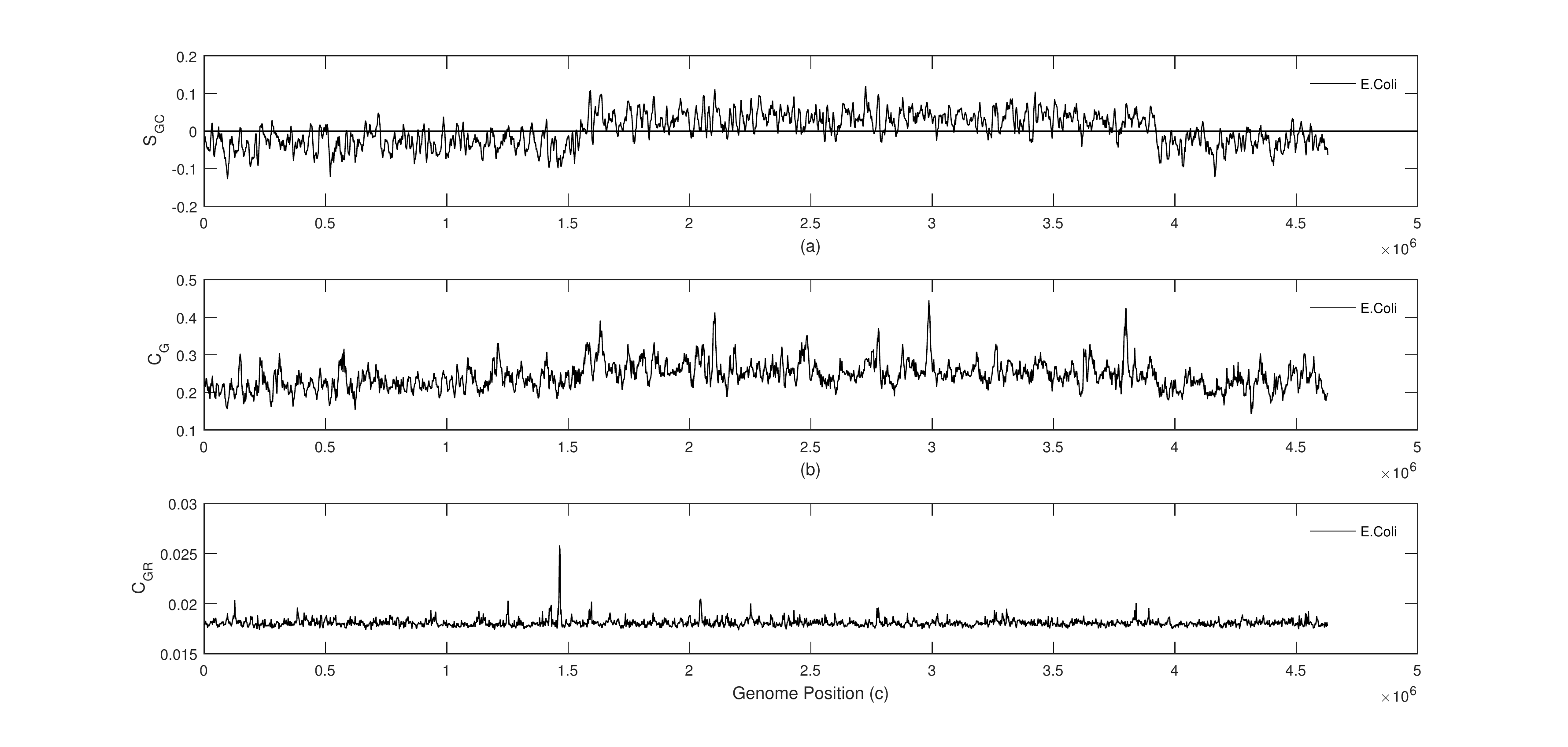} \caption{Plot of the (a) GC skew $\left(S_{GC}\right)$, (b) gCorr $\left(C_{G}\right)$
and (c) iCorr $\left(C_{GR}\right)$ values for \emph{E. coli }str.
K-12 substr. MG1655 (NC\_000913). As can be seen, for \emph{E. coli},
all three methods predict the TER location at nearby genome positions
which also matches with the experimentally known TER location. However,
the graph of iCorr is lot less noisy as compared to GC skew and gCorr,
thereby substantially reducing the ambiguity. The window size is 10000
and shift size is 2000. \label{fig:02}}
\end{figure*}

The auto-correlation method mainly has three limitations. Firstly,
the ORI location is predicted in this method by visually inspecting
the correlation profile which creates room for human error. Secondly,
the window size required in this method is quite large, which becomes a problem for experimental validation. Thirdly, the auto-correlation method uses the positional information of only the
G nucleotide and ignores the statistical properties of other nucleotides. 
In this paper, we propose a modification of this method
 which addresses all these limitations. The proposed complex
correlation method (iCorr) uses four numbers $\left\{ +1,-1,+i=\sqrt{-1},-i\right\} $
and thus is able to represent the positions of each of the four nucleotides,
unlike the auto-correlation method which uses only real numbers $\left\{ +1,-1\right\} $.
In the iCorr method, there is no need for visual inspection and the
ORI/TER region is given by either the location of the peak value of the
computed function. This method requires a much smaller window
size as compared to the auto-correlation method and thus, leads to
a much higher resolution. We have also developed an algorithm to optimize this resolution in order to 
get the best results with a fairly low window size. This mapping of DNA
nucleotides to four unique complex numbers instead of two integers
could also be very effective in solving many other problems of interest in
computational biology \cite{Shah3base,Yau2015,Ruiz2017}.

We describe the iCorr method and the proposed algorithm in Sec. II,
present the results in Sec. III and finally end with conclusions in
Sec. IV.

\section{Methods \label{sec:Methods}}

The primary computational approach for prediction of origin of replication
is to divide the entire genome into overlapping windows/segments of
equal length, and analyse each window to measure some statistical
property using information theory and/or signal processing techniques.
The values thus obtained are plotted against the window number. The
origin of replication is predicted to be present in the window where
a significant change is observed. This abrupt change can manifest
in different ways depending on the actual statistical property being
measured.

\begin{figure*}
\centering{}\includegraphics[clip,width=1\textwidth]{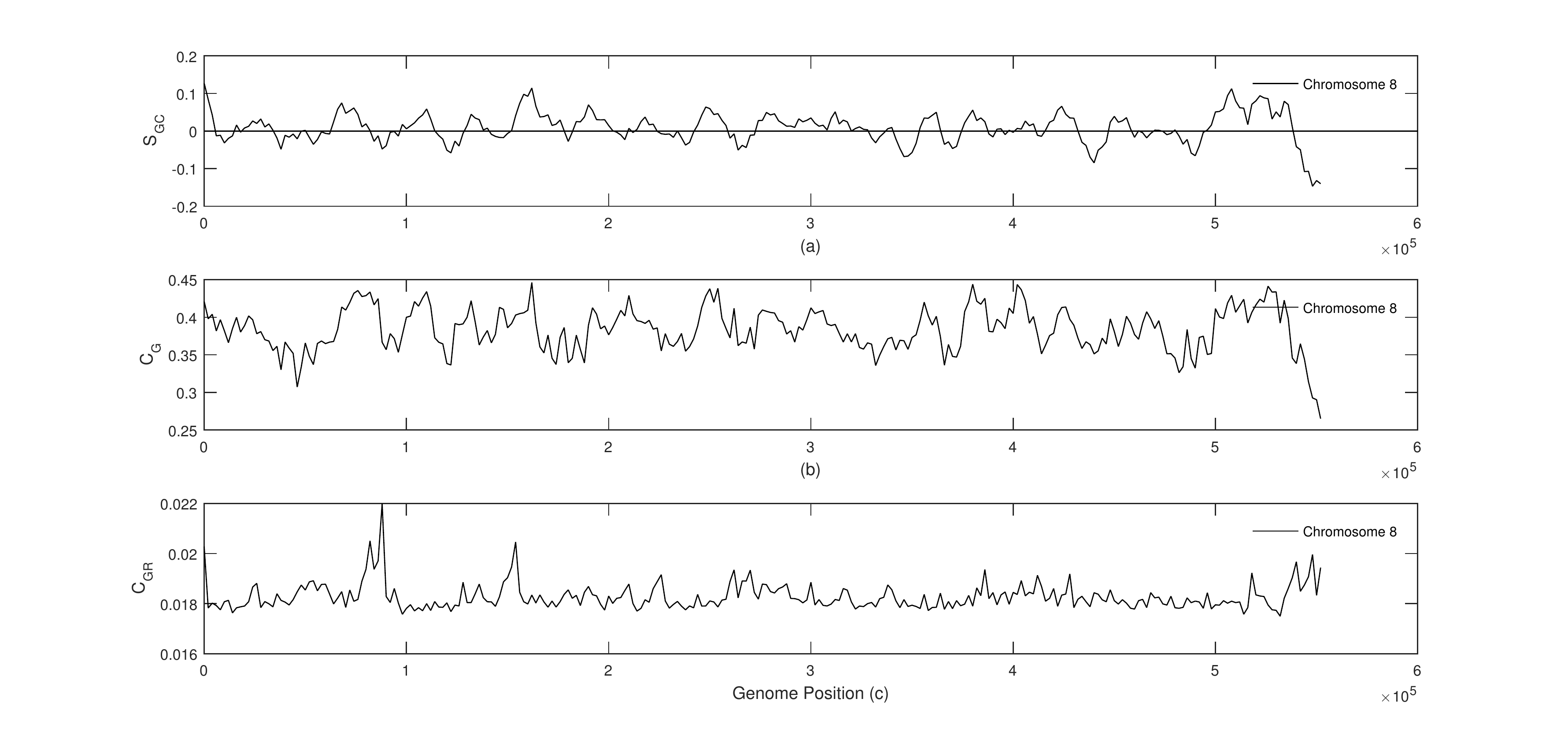} \caption{Plot of the (a) GC skew $\left(S_{GC}\right)$, (b) gCorr $\left(C_{G}\right)$
and (c) iCorr $\left(C_{GR}\right)$ values for chromosome 8 of \emph{S.
cerevisiae}. As can be seen, for this chromosome, the graphs of GC
skew and gCorr are very noisy and do not make any clear predictions.
On the other hand, the iCorr method gives clear peaks which are close
to the experimentally known ORI locations. The window size is 10000
and shift size is 2000.\label{fig:03}}
\end{figure*}

In the gCorr method, the G (Guanine)
nucleotide of each location of the window/segment is denoted by $\left\{ +1\right\} $ and
all other nucleotides by $\left\{ -1\right\} $. This helps in converting
the symbolic sequence to a discrete number sequence thereby making
it conducive for statistical analysis. We calculate the auto-correlation
value of this discrete sequence using the function \cite{Beauchamp1979,Cavicchi2000},
\begin{equation}
C(k)=\frac{1}{(N-k)\sigma^{2}}\sum_{j=1}^{N-k}\left(a_{j}-\mu_{a}\right)\left(a_{j+k}-\mu_{a}\right)\label{eq:C-k}
\end{equation}
where $k=1,2,3,\ldots,N$, $a_{i}\in\left\{ +1,-1\right\} $ denotes
the value at the $i$th position of the discrete sequence, $N$ is
the window size, $\mu_{a}=0$ and $\sigma$= 1 are the means and standard
deviation of the random variable $a_{i}$. The auto-correlation measure,
$C_{G}$, is then defined as the average of all correlation values
in Eq. \eqref{eq:C-k} \cite{Shah2012},

\begin{equation}
C_{G}=\frac{1}{N-1}\sum_{k=1}^{N-1}|C(k)|\label{eq:CG}
\end{equation}
where the subscript ``G'' refers to ``genome''. $C_{G}$ ranges
from 0 to 1 and is independent of the length of the sequence. The
value of $C_{G}$ is a good indicator of the correlation strength
between the positions of the G nucleotide. Thus, a sequence with $C_{G}=0$
corresponds to a lack of correlation and one with $C_{G}=1$ to a
highly correlated sequence.

\begin{figure*}
\centering{}\includegraphics[clip,width=1\textwidth]{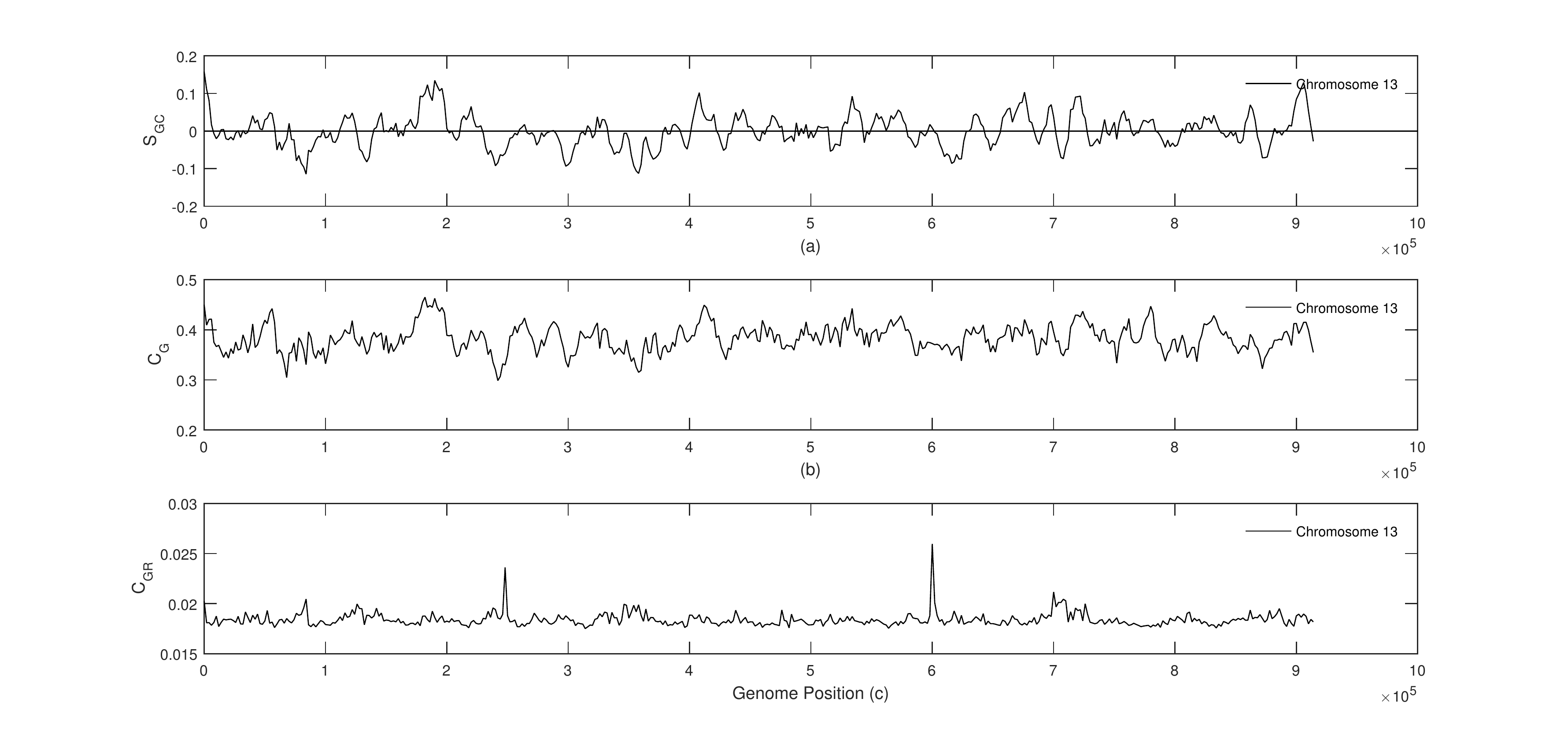} \caption{Plot of the (a) GC skew $\left(S_{GC}\right)$, (b) gCorr $\left(C_{G}\right)$
and (c) iCorr $\left(C_{GR}\right)$ values for chromosome 13 of \emph{S.
cerevisiae}. As can be seen, for this chromosome, the graphs of GC-skew and gCorr are very noisy and do not make any clear predictions.
On the other hand, the iCorr method gives clear peaks which are close
to the experimentally known ORI locations. The window size is 10000
and shift size is 2000.\label{fig:04}}
\end{figure*}

Since a DNA sequence is made up of four bases, we can generate a string
of bits for the
 A (Adenine) base by assigning a value of $\left\{ +1\right\} $
to every occurrence of A and $\left\{ -1\right\} $ to all other positions
(similarly for T and C). In the above method, only the G-track is
chosen for analysis since it gives much better results as compared
to the other three discrete sequences \cite{Shah2012}. Though this
method has been found to work better than the GC-skew method in some
situations, it has an inherent limitation of assigning the same value
of $\left\{ -1\right\} $ to T, A and C. Due to this, it does not
capture the rich variations produced by the four bases present in
DNA sequence.

In this paper, we propose a complex correlation (iCorr) method which
extends the above method to complex states and thereby completely
eliminates the most fundamental limitation in gCorr and other computational
methods for ORI prediction. Furthermore, we propose an algorithm for
automatically optimising the window size and shift size for a given
genome. We use $\{+1,-1,+i=\sqrt{-1},-i\}$ for multi-variate classification
of the four bases present in a DNA sequence. A DNA sequence made up
of ATGC base pairs can give rise to 24 different discrete sequences
using the $\{+1,-1,+i=\sqrt{-1},-i\}$ mapping as opposed to only
4 sequences provided by the $\{+1,-1\}$ mapping. Out of these 24
sequences, only 3 were found to be independent, namely, $\left\{ A\rightarrow+i,\text{ }G\rightarrow-1,\text{ }T\rightarrow-i,\text{ }C\rightarrow+1\right\} $,
$\left\{ A\rightarrow-i,\text{ }G\rightarrow+i,\text{ }T\rightarrow-1,\text{ }C\rightarrow1\right\} $
and $\left\{ A\rightarrow+i,\text{ }G\rightarrow+1,\text{ }T\rightarrow-i,\text{ }C\rightarrow-1\right\} $.
The remaining 21 are equivalent to one of these 3 mappings. After
analysing all these various complex number sequences for ORI prediction,
we have found that the following mapping gives the best results: $\left\{ A\rightarrow+i,\text{ }G\rightarrow+1,\text{ }T\rightarrow-i,\text{ }C\rightarrow-1\right\} $.
The signals obtained from the other 2 choices are very noisy. We calculate
the auto-correlation of this generated discrete sequence using the
same formula given in Eqs. \eqref{eq:C-k} and \eqref{eq:CG} $\left(\text{using }\mu_{a}=0,\text{ }\sigma=1\right)$,
but now $C(k)$ comes out to be a complex number. However, the final
auto-correlation value obtained is still a real number since the RHS
of Eq. \eqref{eq:CG} uses the absolute value (or magnitude) of these
complex $C\left(k\right)$ values. This is the iCorr value (denoted
by $C_{GR}$) and is plotted against the window number (or genome
length). The graph produces a sharp peak at one or more locations.
We propose that the genome position(s) corresponding to these peak
value(s) contain the origin of replication or the termination position
(applicable to prokaryotes which have circular genomes). 

Further, we develop a algorithm with the primary objective of allowing
the user to be able to get the best possible results without the need
to set the window size and shift size beforehand. The algorithm is initially run on 10 different sets of window sizes
and shift sizes, namely: (1000,1000), (2000,1000), (3000,1000), (4000,1000),
(5000,1000), (1000,500), (2000,500), (3000,500), (4000,500) and (5000,500)
to obtain the iCorr curve. For each of the iCorr values obtained,
we perform normalisation of the values obtained given by using the formula:
\begin{equation}
X_{i}=\frac{x_{i}-\mu}{\sigma}\label{eq:C-k-1}
\end{equation}
where $i\in\left\{ 1,2,...,N\right\} $ is the window number, $N$
is the total number of windows, $x_{i}$ is the iCorr value for the
$i$th window, $\mu$ is the mean of all $N$ iCorr values and $\sigma$
is the standard deviation. Out of all the $N$ values of $\left\{ X_{i}\right\} $,
the set $\left\{ Y_{j}\right\} $ contains the values which represent
the peaks corresponding to the replication sites along the genome.
Elements of $\left\{ Y_{j}\right\} $ are all those elements of $\left\{ X_{i}\right\} $
which lie between $X_{L}$ and $X_{H}$, where $X_{H}$ is the maximum
among all $\left\{ X_{i}\right\} $ and $X_{L}$ is calculated as
follows:
\[
X_{L}=\begin{cases}
\text{max}\left(X_{H}\big/3,6\right), & \text{if } X_{H}>10.5\\
6, & \text{if }10.5\ge X_{H}\ge6\\
X_{H}, & \text{if }X_{H}<6
\end{cases}
\]
which implies that if $X_{H}<6$, then $\left\{ Y_{i}\right\} $ is
an empty set and our algorithm cannot predict any replication site
locations along the genome. The numbers above were chosen based on
visual inspection of the resulting data.

We carry out this procedure for all the 10 sets of window sizes and
shift sizes mentioned above, and out of the 10 sets of $\left\{ Y_{i}\right\} $
thus formed containing the topmost peaks, we select one set by comparing
their $X_{H}$ and $X_{L}$ values with each other. If there is one
non-empty set of $\left\{ Y_{i}\right\} $ values with $X_{H}>18$,
we select this one to represent the replication sites. If there are
more than one sets in this category, we select the set with the lowest
value of $X_{H}\big/X_{L}$. This would ensure that the final window
size and shift size values correspond to the graph, all of whose peaks
are possibly distinguisable in terms of amplitude. If there are no
such sets, we then check for non-empty sets with $18\ge X_{H}>10.5$
and if no set is found, then we check for $10.5\ge X_{H}\ge6$. 

\begin{table*}
\begin{longtable*}{|>{\centering}m{2.1cm}|>{\centering}m{2.3cm}|>{\centering}m{10.25cm}|>{\centering}m{2.5cm}|}
\hline 
\textbf{Chromosome Number} & \textbf{Window Size, Shift Size} & \textbf{ORI prediction of iCorr method} & \textbf{OriDB}\tabularnewline
\hline 
\endfirsthead
\hline 
chr01  & 5000,500 & Window Number= 407 Region: 2,03,500-2,08,500 $X_{i}$: 7.51  & Likely\tabularnewline
\hline 
chr02 & 1000,500 & Window Number= 4 Region: 2,000-3,000 $X_{i}$: 29.93 & Confirmed\tabularnewline
\hline 
chr03 & 2000,500 & Window Number= 540 Region: 2,70,000-272000 $X_{i}$: 10.30  & Confirmed\tabularnewline
\hline 
chr04 & 1000,1000 & Window Number= 1309 Region: 13,09,000-13,10,000 $X_{i}$: 10.38

Window Number= 350 Region: 3,50,000-3,51,000 $X_{i}$: 9.67

Window Number= 1188 Region: 11,88,000-11,89,000 $X_{i}$: 8.33

Window Number= 384 Region: 3,84,000-3,85,000 $X_{i}$: 8.08

Window Number= 1427 Region: 14,27,000-14,28,000 $X_{i}$: 6.98

Window Number= 1391 Region: 13,91,000-13,92,000 $X_{i}$: 6.73

Window Number= 442 Region: 4,42,000-4,43,000 $X_{i}$: 6.27 & Likely

Confirmed

Nothing

Dubious

Dubious

Confirmed

Likely \tabularnewline
\hline 
chr05 & 1000,1000 & Window Number= 543 Region: 5,43,000-5,44,000 $X_{i}$: 11.72

Window Number= 2 Region: 2,000-3,000 $X_{i}$: 7.45

Window Number= 171 Region: 1,71,000-1,72,000 $X_{i}$: 7.06 & Nothing

Confirmed

Confirmed \tabularnewline
\hline 
chr06 & 3000,500 & Window Number= 356 Region: 1,78,000-1,81,000 $X_{i}$: 13.21  & Nothing\tabularnewline
\hline 
chr07 & 1000,500 & Window Number= 1504 Region: 7,52,000-7,53,000 $X_{i}$: 10.38

Window Number= 1614 Region: 8,07,000-8,08,000 $X_{i}$: 9.81

Window Number= 1060 Region: 5,30,000-5,31,000 $X_{i}$: 9.01

Window Number= 2178 Region: 10,89,000-10,90,000$X_{i}$: 8.79

Window Number= 2103 Region: 10,51,500-10,52,500 $X_{i}$: 7.13

Window Number= 1442 Region: 7,21,000-7,22,000$X_{i}$: 6.81 & Dubious

Likely

Likely

Confirmed

Nothing

Nothing \tabularnewline
\hline 
chr08 & 1000,1000 & Window Number= 557 Region: 5,57,000-5,58,000 $X_{i}$: 16.93 & Confirmed\tabularnewline
\hline 
chr09 & 4000,500 & Window Number= 128 Region: 64,000-68,000$X_{i}$: 10.13

Window Number= 779 Region: 3,89,500-3,93,500$X_{i}$: 9.43 & Nothing

Nothing\tabularnewline
\hline 
chr10 & 1000,1000 & Window Number= 715 Region: 7,15,000-7,16,000 $X_{i}$: 10.64

Window Number= 3 Region: 3,000-4,000 $X_{i}$: 7.96 & Confirmed

Confirmed\tabularnewline
\hline 
chr11 & 5000,1000 & Window Number= 621 Region: 6,21,000-6,26,000 $X_{i}$: 14.25 & Dubious\tabularnewline
\hline 
chr12 & 3000,1000 & Window Number= 750 Region: 7,50,000-7,53,000$X_{i}$: 10.63

Window Number= 304 Region: 3,04,000-3,07,000 $X_{i}$: 9.72

Window Number= 104 Region: 1,04,000-1,07,000$X_{i}$: 6.79

Window Number= 123 Region: 1,23,000-1,26,000 $X_{i}$: 6.36 & Dubious

Nothing

Nothing

Nothing \tabularnewline
\newpage
\hline 
chr13 & 1000,1000 & Window Number= 410 Region: 4,10,000-4,11,000 $X_{i}$: 11.37

Window Number= 610 Region: 6,10,000-6,11,000 $X_{i}$: 10.45

Window Number= 85 Region: 85,000-86,000 $X_{i}$: 7.24 & Nothing

Confirmed

Nothing \tabularnewline
\hline 
chr14 & 1000,1000  & Window Number= 455 Region: 4,55,000-4,56,000$X_{i}$: 10.50

Window Number= 55 Region: 55,000-56,000 $X_{i}$: 7.48

Window Number= 705 Region: 7,05,000-7,06,000 $X_{i}$: 7.19 & Dubious

Nothing 

Dubious\tabularnewline
\hline 
chr15 & 3000,1000  & Window Number= 30 Region: 30,000-33,000 $X_{i}$ : 11.80 

Window Number= 345 Region: 3,45,000-3,48,000 $X_{i}$: 11.21 

Window Number= 895 Region: 8,95,000-8,98,000 $X_{i}$: 8.29 

Window Number= 236 Region: 2,36,000-2,39,000 $X_{i}$: 7.14 

Window Number= 713 Region: 7,13,000-7,16,000 $X_{i}$: 6.39  & Confirmed

Likely

Likely

Dubious

Nothing \tabularnewline
\hline 
chr16 & 2000,1000 & Window Number= 818 Region: 8,18,000-8,20,000 $X_{i}$: 11.83 

Window Number= 522 Region: 5,22,000-5,24,000 $X_{i}$: 8.51  & Confirmed

Nothing \tabularnewline
\hline 
\caption{Results of iCorr optimisation algorithm applied to 16 chromosomes of \emph{S.
cerevisiae} and its comparison with experimental results. \label{tab:Results-of-optimisation}}
\end{longtable*}
\end{table*}

\section{Results\label{sec:Results}}

We have applied the method described in the previous section to two
bacterial genomes obtained from NCBI \cite{NCBI2016} and 16 chromosomes
of one eukaryote (\emph{S. cerevisiae}) obtained from OriDB \cite{Siow2012}.
In this section, we describe the results obtained.

Figures \ref{fig:01} and \ref{fig:02} show a plot of the (a) GC
skew $\left(S_{GC}\right)$, (b) gCorr $\left(C_{G}\right)$ and (c)
iCorr $\left(C_{GR}\right)$ values for two prokaryotic genomes, \emph{B.
subtilis} and \emph{E. coli}, respectively. In GC skew method, the
ORI (or TER) location is given by the point where the graph crosses
the zero line from above (or below). In gCorr method, the ORI/TER
location is given by the position where the graph undergoes a sharp
jump (higher or lower). In iCorr method, the ORI/TER location is given
by the position of peak values. As can be clearly seen, all these
three methods correctly predict the TER location for \emph{B. subtilis}
and \emph{E. coli}. However, the graphs for GC skew and gCorr are
lot more noisy as compared to the graph for the iCorr method. In Fig.
\ref{fig:01}a and \ref{fig:02}a, there are also several other points
of zero-crossing which can be erroneously considered to the ORI/TER
location thereby making the GC skew prediction quite ambiguous. And
this problem only becomes worse as we further reduce the window size.
The gCorr method predicts the presence of ORI/TER in a genome where a
sudden transition is observed. The transition spans several windows
and its detection depends on human judgement which reduces the accuracy
in ORI prediction. In contrast, the iCorr method for prokaryotes predicts
the location by finding peak in the graph. Peak is obtained at a single
point which helps to narrow down our area of interest to a single
window. In the case of \emph{B. subtilis}, the gCorr predicts the
ORI to be present in a genome segment whose length is around 100k
nucleotides (see Fig.\ref{fig:01}b). In contrast, the iCorr method
can bring down the range to 10k nucleotides which implies a 10 times
higher resolution.

Compared to prokaryotic genomes, the computational prediction of ORI
in eukaryotic genomes has been considerably much more challenging
due to the rich and complex structure of DNA with multiple ORI being
present in a single chromosome. And an added disadvantage is that
experimentally verified ORI locations are available for only a few
eukaryotes like \emph{S. cerevisiae}. Figures \ref{fig:03} and \ref{fig:04}
show the graph of the various methods for chromosome 8 and 13 of S.
\emph{cerevisiae} respectively. As can be seen, the plot of the GC
skew and gCorr method in Figs. \ref{fig:03} and \ref{fig:04} is
very noisy and thus, does not help in making any clear prediction. On the other
hand, the iCorr method gives clear peaks which are actually close
to the experimentally known ORI locations. Note that, unlike the prokaryotic
genome which are circular, eukaryotic genomes are linear and do not
have a separate TER region. Hence, its not clear whether all zeros
of the GC skew plot correspond to ORI or only those where the graph
crosses zero from above. 

Details of the ORI prediction by iCorr method for all chromosomes
of \emph{S. cerevisiae} and the closest experimentally confirmed ORI
are given in Table \ref{tab:Results-of-optimisation}. Column 1 of
this table denotes the chromosome number. Column 2 denotes the optimized
window size and shift size given by our algorithm described in the
previous section. Column 3 denotes the window number and region along
the chromosome at which a peak was detected, and also the normalised
value of the peak. Column 4 denotes whether that detected peak corresponds
to a Confirmed, Likely or Dubious, experimental detection of ORI as
given in the OriDB database \cite{Siow2012}. Mention of Nothing in
this column denotes that this genome location does not correspond
to any experimentally detected ORI location. To compare our predictions
with experiments, we took a region flanked by 5000 base pairs on both
sides of our computational prediction. As can be seen in this table,
the algorithm was able to predict at least one experimentally detected
ORI location (Confirmed or Likely) for 11 out of the 16 chromosomes. 

It is important to note that our iCorr method predicts only a
few ORI locations in \emph{S. cerevisiae} chromosomes which actually
contain many experimentally verified ORI locations \cite{Siow2012}. This clearly indicates
that different ORI locations in eukaryotes have different statistical
properties which might require more sophisticated computational methods
for their correct identification. Also, not all ORI locations may
be predictable by using the same statistical measure. This holds for
prokaryotes as well, where the ORI locations of different genomes can
have different statistical properties, thereby requiring different
computational tools for ORI prediction.

\section{Discussion\label{sec:Discussion}}

In the past, several methods have been developed to predict ORI location
for prokaryotes but most of them utilised only a limited amount of
information present in the DNA sequence. The GC skew method \cite{Mrazek1998}
considered frequency counts of G and C nucleotides as the sole means
to predict ORI location and neglected the importance of positioning
of each base in a DNA sequence. The auto-correlation based gCorr method
was developed to remove this inherent flaw of GC skew method by considering
relative base positions of the G nucleotide. However, this method
was unable to differentiate between A, C and T nucleotides. In an
attempt to fully discover the rich variety of bases present in a sequence,
we have extended the basic gCorr method to complex states. The iCorr
method presented in this paper takes into consideration the relative
base positioning of all the four nucleotides. This method has been
found to significantly improve the resolution of ORI prediction of
prokaryotes and has also been able to predict the ORI locations of
\emph{S. cerevisiae} to a good extent. The prediction of iCorr method currently does not match with the experimentally verified ORI locations for \emph{P. falciparum} \cite{Agarwal2017}, but we hope that we will be able
to validate and refine our methods as more experimental data becomes
available in the future for this and other genomes.

Similar to all the previously existing computational methods, iCorr
only suggests the ORI/TER location and does not guarantee its existence,
which needs to be experimentally verified. With the advantages of
pin-point peak detection and utilisation of rich structure present
in DNA, the iCorr method is a significant progress in ORI prediction
for prokaryotes and eukaryotes. Here it is important to note that
the predictions made by these computational methods are significantly
dependent on the choice of window/segment size into which the genome
is divided for statistical analysis. If the window size is taken to
be too large, then the meaningfulness of the predictions obviously
goes down. And if the window size is taken to be too small, the graphs
can be very noise and lead to decrease in accuracy and precision.
Optimization of the window size and shift size for a given genome
is an open problem which we have tried to tackle in this paper with the help
of our proposed algorithm.

\begin{acknowledgments}
We would like to thank Ayush Jain for helping us with some part of
the analysis. A part of this work was done while KS was at the Indian
Institute of Technology (IIT) Delhi. 
\end{acknowledgments}

\end{document}